\documentstyle[prd,aps]{revtex}

\begin{document}
\tightenlines
%\draft
\parskip 0.3cm
\begin{titlepage}

\begin{centering}
{\large \bf $\tau$ Anomalous Couplings and Radiation Zeros in the
$e^+e^- \rightarrow \tau \bar\tau \gamma$ process}\\ \vspace{.9cm}
{\bf I.Rodriguez and O.A.Sampayo}\\
\vspace{.05in}
{\it  Departamento de F\'{\i}sica,
Universidad Nacional de Mar del Plata \\
Funes 3350, (7600) Mar del Plata, Argentina} \\ \vspace{.4cm}
\vspace{.08in}
{ \bf Abstract}
\\
\bigskip
\end{centering}
{\small The process $e^+e^- \rightarrow \tau \bar\tau \gamma$
contains configurations of the four-momenta for which the
scattering amplitude vanishes (Radiation Zeros or Null Zone).
These Radiation Zeros only occur for couplings given by a gauge
theory, in particular the standard model. Therefore they are
sensitive to physics beyond the standard model as the anomalous
magnetic and weak moment of the $\tau$ lepton. In this article we
compute the effect of these anomalous interactions on the
Radiation Zeros in the above mentioned process.}


\vspace{0.2in}
\vfill

%\begin{flushleft}
%MDP-HEP/98-01\\
%October 1998 \hfill
%\draft
%\end{flushleft}
\end{titlepage}
\vfill\eject

\section{\bf Introduction}

The bounds on the anomalous magnetic moment of the $\tau$ lepton
$a_{\gamma}$, are much weaker than the ones for electron and muon.
Due to the $\tau$ lepton's short lifetime of $(291.0\pm1.5)\times
10^{-15}$ seg, its anomalous magnetic moment can not be measured
by a spin precession method and no direct measurement of
$a_{\gamma}$ exists so far. Because of the impossibility of
measuring $a_{\gamma}$ by a spin precession method, the present
bounds have been obtained by analysis of collision experiments. In
that sense, interesting articles dealing with the computation of
bounds for $a_{\gamma}$ have been recently published. The OPAL
Collaboration \cite{opal} uses a reaction proposed by A. Mendez
and A. Grifolds \cite{mendez} some years ago. They obtained limits
on $a_{\gamma}$ from the non-observation of anomalous $\tau
\bar{\tau} \gamma$ production at LEP. In other article,
G.A.Gonzales-Sprinberg, A.Santamaria and J.Vidal using LEP1, SLD,
and LEP2 data for tau lepton production, and data from CDF, D0 and
LEP2, for W-decays into tau lepton, established model independent
limits on non-standard electromagnetic and weak magnetic moments
of the tau lepton \cite{gabriel}. These obtained bounds are still
far away from the theoretical precision for $a_{\gamma}$:$(1177.3
\pm 0.3)\times 10^{-6}$. This weakness of the $a_{\gamma}$ bounds
is unfortunate since large deviations from the S.M. values are
expected for the $\tau$ lepton. In particular, in composite models
one would expect larger effects for the tau lepton than for the
rest of the leptons.

In this article we are interested in studying the effect of the
anomalous magnetic and weak moment of the $\tau$ lepton on
radiation zeros. In this respect we study the Radiation Zeros to
arise in the $e^+e^- \rightarrow \tau \bar\tau \gamma$ process at
future $e^+e^-$ collider energies. These kind of zeros are know as
type 2 zeros \cite{ceros}. Definitions and study of this kind of
zeros is presented in section 2 in the soft photon approximation.
This discussion should be viewed as a qualitative argument for the
existence and the location of radiation zeros. Our purpose there
is only to show, in a very simple fashion, where the radiation
zeros are. When we numerically calculate the cross-section using a
Monte Carlo technique in section 4 we do not make use of this
approximation. There we calculate the cross section without
approximation on the photon energy and we only impose cuts on the
photon energy to ensure that the photon is the softest particle in
the final state. In this conditions the numerical results for the
radiation zeros positions should be similar to obtained in the
soft photon approximation.

Following Ref \cite{masso}, in order to analyze the tau magnetic
and weak moments, we will use an effective lagrangian description.
Thus, in section 3, we describe the effective lagrangian
formalism.  In section 4 we present the results for the zeros
position, the standard model  cross section and the effective
coupling effects. Finally, in section 5 we give our conclusions.

\section{\bf Radiation Zeros}

In certain high-energy scattering processes involving charged
particles and the emission of one photon there are configurations
of the final state particles for which the scattering amplitude
vanishes. Such null zones are known as radiation zeros. Detailed
study about this effect can be found in the literature for several
process \cite{ceros}. The process that we are studying in this
paper ($e^+e^- \rightarrow \tau \bar\tau \gamma$) presents
radiation zeros of type 2. For this kind of zeros is not necessary
that the charged particles have same sing charges but all
particles including the photon have to lie in the same plane. To
have a first idea about the location of these zeros we present in
this section a study within the soft photon approximation. The
discussion is particularly simple when one considers the amplitude
for the emission of a single soft photon.

The matrix elements for one soft photon emission can be written as

\begin{eqnarray}
M_{\gamma}\propto J_{\mu} \epsilon^{\mu} M_0
\end{eqnarray}

where $M_0$ is the leading order 4-fermion matrix element and
$\epsilon^{\mu}$ is the photon polarization. In the other hand the
current $J_{\mu}$ is given by

\begin{eqnarray}
J_{\mu}=\sum_i \alpha_i \frac{p^i_{\mu}}{p^i.k}
\end{eqnarray}

where the sum run over the four charged fermions ($e,\bar e, \tau,
\bar\tau$) with 4-momenta $p_1$, $p_2$, $p_3$ and $p_4$
respectively and $k$ is the photon 4-momenta. The coefficient
$\alpha_i$ is $-1$ for $e$ and $\bar\tau$ and $1$ for $\tau$ and
$\bar e$.

In the soft photon approximation with massless fermions, we have
$k^0=\omega_{\gamma}$, $p_1^0=p_2^0=p_3^0=p_4^0=\sqrt{s}/2$. In
this conditions, if we denote by $\check{n_i}$ the directions for
the spatial momentum of the charged fermions and by $\check{n}$
the direction for the photon spatial momentum , then we have
$\overrightarrow{p_i}=\frac{\sqrt{s}}{2} \check{n_i}$ and
$\overrightarrow{k}=\omega_{\gamma} \check{n}$.

Using the 4-momentum conservation in the center of mass frame

\begin{eqnarray}
\frac{\sqrt{s}}{2}\check{n_3}+\frac{\sqrt{s}}{2}\check{n_4}+\omega_{\gamma}
\check{n}=0
\end{eqnarray}

 we have the obvious condition that we call {\bf condition 1} :
The 3-momentum of the three final state particles (directions
$\check{n_3}$, $\check{n_4}$ and $\check{n}$ ) lie in the same
plane (Moreover, in the soft photon approximation where
$\omega_{\gamma} \rightarrow 0$, we have
$\check{n_4}=-\check{n_3}$).

On the other hand the radiation zero condition implies that

\begin{eqnarray}
M=0 \mbox{\hspace{1cm} thus \hspace{1cm}} J^{\mu}\epsilon_{\mu}=0
\end{eqnarray}

Considering the expression for the current $J_{\mu}$ we have,
according to gauge invariance, that

\begin{equation}
J^{\mu}k_{\mu}=\sum_{i=1,4}\alpha_i \frac{p_i^{\mu}}{p_i.k} k_{\mu}=\sum_{i=1,4}
\alpha_i=0
\end{equation}

In the other hand by gauge invariance we have that
$\epsilon_{\mu}k^{\mu}=0$. In this conditions the transformation
\begin{eqnarray}
J^{\mu} &\rightarrow& J^{\mu} + a k^{\mu} \\
\epsilon^{\mu} &\rightarrow& \epsilon^{\mu} + b k^{\mu}
\end{eqnarray}
with arbitrary $a$ and $b$ leaves $M$ invariant.

Then we can choose the numbers $a$ and $b$ such that
$\epsilon^{\mu}=(0,\overrightarrow{\epsilon})$ and
$J^{\mu}=(0,\overrightarrow{J})$.

In this conditions, the Zero Radiation request implies
$\overrightarrow{J}.\overrightarrow{\epsilon}=0$, but if we impose
this condition for every photon polarization then the useful
conditions read $\overrightarrow{J}=0$.

To find the expression for $\overrightarrow{J}$ we have first to
find the constant $a$ impossing that $J^0=0$, then
\begin{equation}
J^0=\sum_{i=1,4} \alpha_i \frac{p_i^0}{p_i.k}  +  a k^0 = 0
\end{equation}

If we call $x_i=\cos\theta_{i,\gamma}$, where $\theta_{i,\gamma}$ is the angle
beteween the directions of the photon and the fermion $i$, then
\begin{equation}
a=\frac{2}{\omega_{\gamma}^2}\left(\frac{x_1}{1-x_1^2}-\frac{x_3}{1-x_3^2}
\right),
\end{equation}
where we have used $k^0=\omega_{\gamma}$, $p_i^0=\sqrt{s}/2$,
$x_2=-x_1$ and $x_4=-x_3$ which is valid in the soft photon
approximation.

With the $a$ value found above we obtain the following expression
for $\overrightarrow{J}$
\begin{equation}
\overrightarrow{J}=\frac{2}{\omega}\left[\frac{-1}{1-x_1^2}\check{n_1}+
\frac{1}{1-x_3^2}\check{n_3}+\left( \frac{x_1}{1-x_1^2}-\frac{x_3}{1-x_3^2}\right)
\check{n}\right]
\end{equation}
which obviously satisfies $\overrightarrow{J}.\check{n}=0$ (gauge
invariance).

By using the Radiation Zero condition $\overrightarrow{J}=0$ we
have
\begin{equation}
\frac{-1}{1-x_1^2}\check{n_1}+
\frac{1}{1-x_3^2}\check{n_3}+\left( \frac{x_1}{1-x_1^2}-\frac{x_3}{1-x_3^2}\right)
\check{n}=0
\end{equation}
which implies that the vectors $\check{n_1}$, $\check{n_3}$ and
$\check{n}$ lie in same plane. That is we call the {\bf condition
2}. Thus taking into account both conditions ({\bf 1} and {\bf
2}), we obtain the necessary condition for radiation zeros: the
momenta of all the particles included the photon lie in the same
plane. We call this the {\it planarity condition}. In order to
find the exact position of the radiation zeros in the soft photon
approximation, we proyect the vectorial equation (11) over the
directions $\check{n_1}$ and $\check{n_3}$. Calling
$x_{31}=\cos\theta_{31}$, we obtain the equations
\begin{eqnarray}
x_{31}-x_1x_3=1-x_3^2 \\
x_{31}-x_1x_3=1-x_1^2
\end{eqnarray}

then $x_3=\pm x_1$. Using the solution $x_3=x_1$ in some of the
above equations we obtain $x_{31}=1$. This solution is
unintersting as all particles are in the beam axes. Considering
now the solution $x_3=-x_1$ and replacing it in the equations (12)
or (13), we find a solution for the positions of the radiation
zeros:
\begin{equation}
x_1=\pm\frac{1}{\sqrt{2}}\sqrt{1-x_{31}},
\end{equation}

or in function of the angles shown in Fig. 2 we obtain according
to Ref.\cite{ceros}

\begin{equation}
\theta_\gamma=\arccos \left[\pm
\frac{1}{\sqrt{2}}\sqrt{1-\cos\theta_{CM}} \right]
\end{equation}

 The above equation
give us the radiation zeros positions. There are two zeros for
each value of the angle between the outgoing $\tau$ and the beam
axes, as we show in fig 3.

\section{\bf The effective lagrangian approach}

In the last few years, effective lagrangians have been used as an
adequate framework to study low energy effects of physics beyond
the standard model (SM). Since the SM gives a very good
description of all physics at the energies available at present
accelerators, then one expects that any deviation of the SM can be
parametrized by an effective lagrangian built with the same fields
and symmetries that the SM. In this conditions, the effective
lagrangian contains a renormalizable piece, the SM theory, and
non-renormalizable operators of dimension higher than 4, which are
suppresed by inverse powers of the high energy physics scale,
$\Lambda$. The leading non-standard effects will come from the
operators with the lowest dimension. Those are dimension six
operators. In particular, there are only two six-dimension
operators which contribute to the anomalous magnetic moments
\cite{buchmuller}:

\begin{eqnarray}
{\cal O}_B&=&\frac{g'}{2 \Lambda^2} \bar L_L \phi \sigma_{\mu\nu}
\tau_R B^{\mu\nu} \\ {\cal O}_W&=&\frac{g}{2 \Lambda^2} \bar L_L
\overrightarrow{\tau} \phi \sigma_{\mu\nu} \tau_R
\overrightarrow{W}^{\mu\nu} \nonumber
\end{eqnarray}

where $L_L=(\nu_L,\tau_L)$ is the tau leptonic doublet and $\phi$
is the Higgs doublet. $B^{\mu\nu}$ and $W^{\mu\nu}$ are the
$U(1)_Y$ and $SU(2)_L$ field strength tensors, and $g'$ and $g$
are the corresponding gauge couplings. Thus, we write our
effective Lagrangian as

\begin{eqnarray}
{\cal L}_{eff} = {\cal L}_{SM} + \alpha_B {\cal O}_B + \alpha_W
{\cal O}_W + h.c.
\end{eqnarray}

As we are not interested in studying $CP$ violation effects, then
we will take the coupling $\alpha_B$ and $\alpha_W$ as real.
Moreover, we will consider them as free parameters without any
further assumption.

The interaction lagrangian can be written in terms of the physical
fields $A_{\mu}$, $Z_{\mu}$ and $W^{\pm}_{\mu}$. In our particular
case, we are only interested in the effective electromagnetic and
 neutral weak interaction, since we are studying a process which
 involves only interactions with $\gamma$ and $Z$ bosons. In this
conditions, the relevant lagrangians is

\begin{eqnarray}
{\cal L}_{eff}={\cal L}_{SM}+\epsilon_{\gamma} \frac{e}{2 m_{Z}}
\bar \tau \sigma_{\mu\nu} \tau F^{\mu\nu} + \epsilon_{Z}
\frac{e}{2 m_{Z}\sin{\theta_W}\cos{\theta_W}} \bar \tau
\sigma_{\mu\nu} \tau Z^{\mu\nu}+\cdots
\end{eqnarray}

where the dots represent non-relevant pieces of the lagrangian and
$F_{\mu\nu}$ and $Z_{\mu\nu}$ are the electromagnetic and weak
field strength tensor respectively. We have expressed the coupling
in function of the parameters $\epsilon_{\gamma}$ and
$\epsilon_Z$. This parameters are functions of the constants
$a_{\gamma}$ and $a_Z$ which are directly related to the
experimental measurement and theoretical calculations:

\begin{eqnarray}
\epsilon_{\gamma}&=& \frac{m_Z}{2m_{\tau}} a_{\gamma} \\
\epsilon_{Z}&=& \frac{m_Z \sin{\theta_W}\cos{\theta_W}}{2m_{\tau}}
a_{Z}
\\
\end{eqnarray}

%%%%%%%%%%%%%%%%%%%%%%%%%%%%%%%%%%%%%%%%%%%%%%%%%%%%%%%%%%%%%%%%%%%%%%%%%%%
\section{\bf  The $e^+e^- \rightarrow \tau \bar \tau \gamma$ process}

In this section we study the $e^+e^- \rightarrow \tau
\bar\tau\gamma$ process which involves electro-weak interactions
plus additional anomalous magnetic and weak moment couplings given
by the operators in Eq. (16). The corresponding Feynman diagrams
are shown in Fig.1. We do not present the amplitude explicitly
here because it is a very long and not illuminating expression.
The calculation was done in the helicity formalism \cite{heli} and
the mass of the final leptons was considered as null due to the
high center of mass energies considered. The blak blob in Fig.1
represents the effective operator contributions. The kinematics of
the process is represented in Fig. 2 where the relevant scattering
angles are shown. The intensity of the effective vertices are
measured by the parameters $\epsilon_{\gamma,Z}$. We take
different and representative values of typical bounds founded by
another authors \cite{opal,mendez,gabriel}. To have a first
filling of the effective vertices effects on the radiation zeros
we plot in Fig. 4 the square amplitude as a function of
$\theta_{\gamma}$ for $\theta_{CM}=20^0$ and $\sqrt{s}=1000$ GeV.
We include (dotted line) the effective contribution for
$\epsilon_{\gamma}=\epsilon_{Z}=0.1$. This figure does not
represent any measurable quantities but it is included here to
show the effect of the anomalous couplings on the radiation zeros
in the squared scattering amplitude. As we can see, the effective
contribution destroyes the standard model radiation zero and then
became itself in a possible observable to bound anomalous effects
over the $\tau$ interactions. However, in practice, experiments
deal with binned quantities and a more realistic study should take
these into account.  In this conditions, we have done a Monte
Carlo calculation of the cross section, where we have included
realistic cuts to eliminate the collinear singularity and to
impose the planarity condition for type 2 radiation zeros. The
numerical integration was done using the numerical subroutine
RAMBO \cite{rambo} and considering three different center of mass
energies and  sets of cuts adequate for each one. The center of
mass energies considered were 500, 1000 and 2000 GeV. The aim is
to see whether the radiation zeros remain visible after a more
realistic analysis. We generate a sample of $\bar \tau \tau
\gamma$ events using a Monte Carlo which includes the exact phase
space and matrix elements without approximation on the photon
energy. The following sequence of cuts is applied for each center
of mass energies:
\begin{eqnarray}
\sqrt{s}&=&500\hspace{1mm}GeV \mbox{\hspace{.5cm}} \rightarrow
\mbox{\hspace{.5cm}}
10GeV<\omega_{\gamma}<100GeV<E_{\bar\tau},E_{\tau}  \nonumber
\\ \sqrt{s}&=&1000\hspace{1mm}GeV \mbox{\hspace{.5cm}} \rightarrow
\mbox{\hspace{.5cm}}
10GeV<\omega_{\gamma}<200GeV<E_{\bar\tau},E_{\tau}    \\
\sqrt{s}&=&2000\hspace{1mm}GeV \mbox{\hspace{.5cm}} \rightarrow
\mbox{\hspace{.5cm}}
10GeV<\omega_{\gamma}<400GeV<E_{\bar\tau},E_{\tau}    \nonumber
\end{eqnarray}

These cuts ensure that the photon is the softest particle in the
final state and then the positions of the radiation zeros are near
to the ones obtained in the soft photon approximation. (Moreover
it is required that the photon must be separated in angle from the
beam and $\tau$ and $\bar \tau$ directions).
\begin{equation}
\theta_{\gamma,beam} > 5^0, \mbox{\hspace{.5cm}}
\theta_{\gamma,\tau}>2^0, \mbox{\hspace{.5cm}},
\theta_{\gamma,\bar\tau}>2^0
\end{equation}
With this cuts we define a measurable sample of $\tau \bar \tau
\gamma$ events. In order that the studied process shows radiation
zeros we have to impose the planarity condition. This condition is
achieved by requiring that the normal to the plane defined by the
beam and the outgoing $\tau$ and the plane defined by the photon
and the $\bar \tau$ directions are approximately parallel. With
the notation of Fig 2 this condition reads  \cite{ceros}:

\begin{equation}
|\breve{n}_{13}.\breve{n}_{1k}|>\cos20^0,
\end{equation}

where $\breve{n}_{13}$ and $\breve{n}_{1k}$ are the normal vector
to the plane containing the initial $e^-$ and final $\tau$ and the
initial $e^-$ and the final $\gamma$ respectively. For the
direction of the lepton $\tau$ we have considered a bin centered
in $\theta_{\tau}=\theta_{CM}$ of width $10^0$. In this conditions
we integrate over

\begin{equation}
\theta_{CM}-5^0 < \theta_{\tau} < \theta_{CM}+5^0
\end{equation}

Although we choose cuts that mimic the soft-photon kinematics we
do not expect strict zeros due to the binnig, the planarity cut
and the integration over the photon energy. In this conditions,
close to the zeros positions as indicated in Fig. 3 , we expect
dips in the photon distribution.

In Figs. 5-10 we show the $\theta_{\gamma}$ distribution for
different values of $\theta_{CM}$ and for different values of the
center of mass energy. In this plots we include the contribution
from the effective interactions for different values of
$\epsilon=\epsilon_{\gamma}=\epsilon_{Z}$ . The results of this
approximated calculation shows that the effective contribution
does not exhibit dips of the same depth as the standard model one.
The effect is small at low energy but due the behaviour of the
effective coupling the effect grows appreciably with the center of
mass energy. The aim of the present work is only show how the
anomalous tau lepton couplings destroy the radiation zeros and
produce, in principle, an observable effect. Study about the real
sensitivity of this effect require a more realistic analysis with
realistic detection errors which is beyond the scope of the
present work.

\section{Conclution}

In this work we have investigated the possibility to use radiation
zeros in the $e^+e^- \rightarrow \tau \bar\tau \gamma$ process to
bound anomalous contributions to $\tau$ magnetic and weak moment.
We have implemented a Monte Carlo program to integrate the phase
space including realistic cuts. The anomalous contribution is seen
as an apartment from the standard model prediction in the regions
where it has radiation zeros. To decide if this kind of observable
could be useful it is necessary a most realistic study that takes
account of the possible stage of high luminosity in the
electron-positron collider.

{\bf Acknowledgements}

We thank CONICET (Argentina), Universidad Nacional de Mar del
Plata (Argentina) for their financial supports.

\pagebreak

\noindent{\large \bf Figure Captions}\\

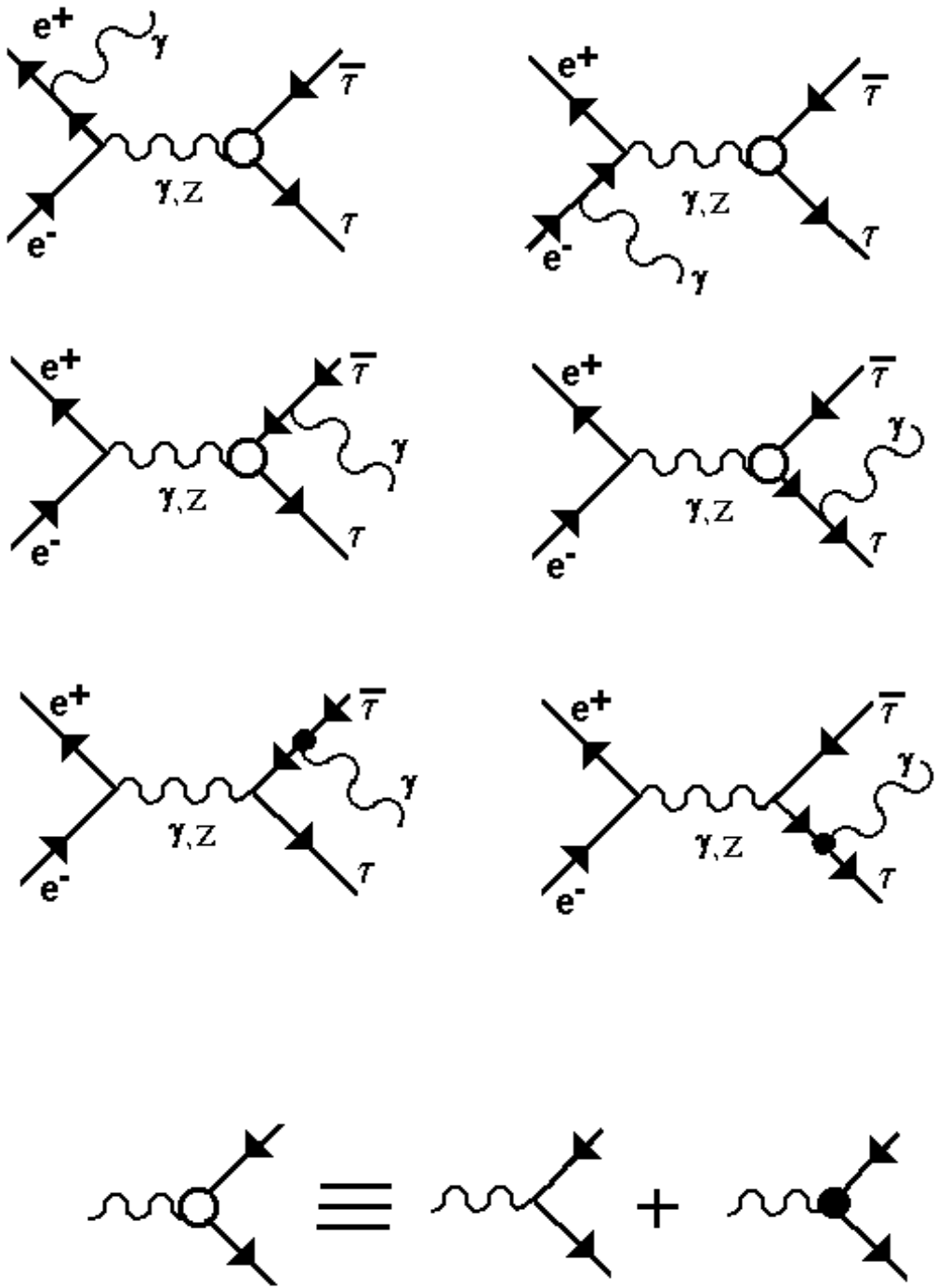
\noindent{\bf Figure 1:} Feynman graph contributing to the
amplitude of the $e^+e^- \rightarrow \tau \bar\tau \gamma$
process.

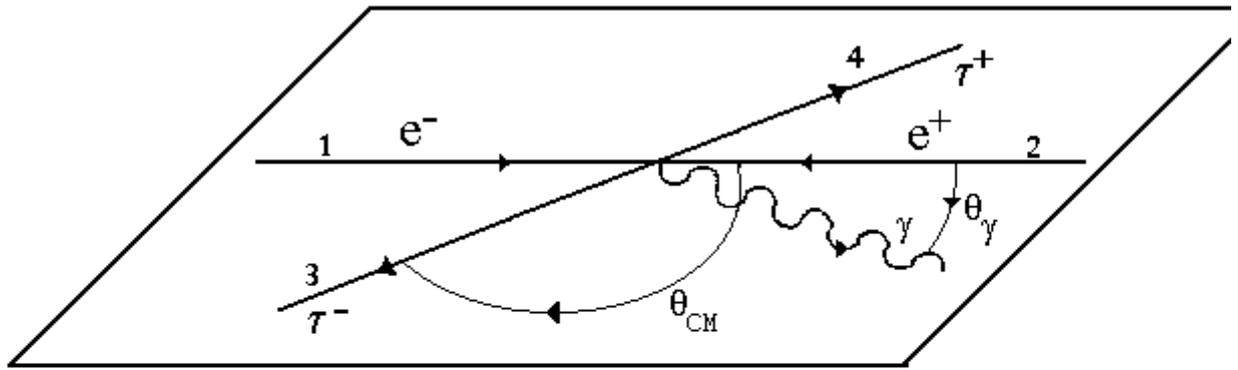
\noindent{\bf Figure 2:} Parameterisation of the kinematics for
the soft photon  $e^+e^- \rightarrow \tau \bar\tau \gamma$
scattering in the $e^+e^-$ center of mass frame.

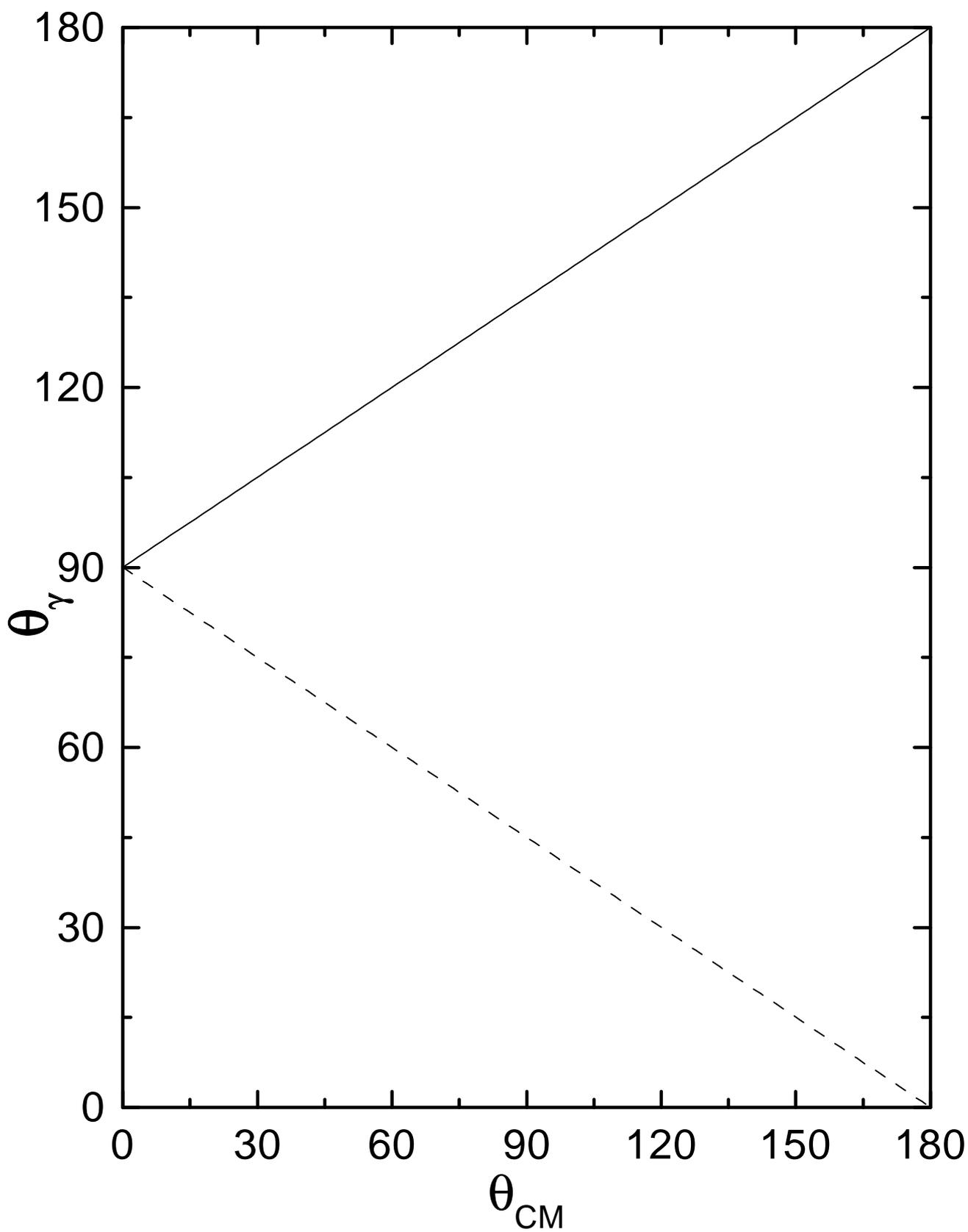
\noindent{\bf Figure 3:} Radiation zero position for $e^+e^-
\rightarrow \tau \bar\tau \gamma$ scattering in the
$\theta_{CM}$,$\theta_{\gamma}$ plane. The solid (dashed) line
represents
$\theta_{\gamma}=\arccos[\mp\frac1{\sqrt{2}}\sqrt{1-\cos\theta_{CM}}]$
respectively.

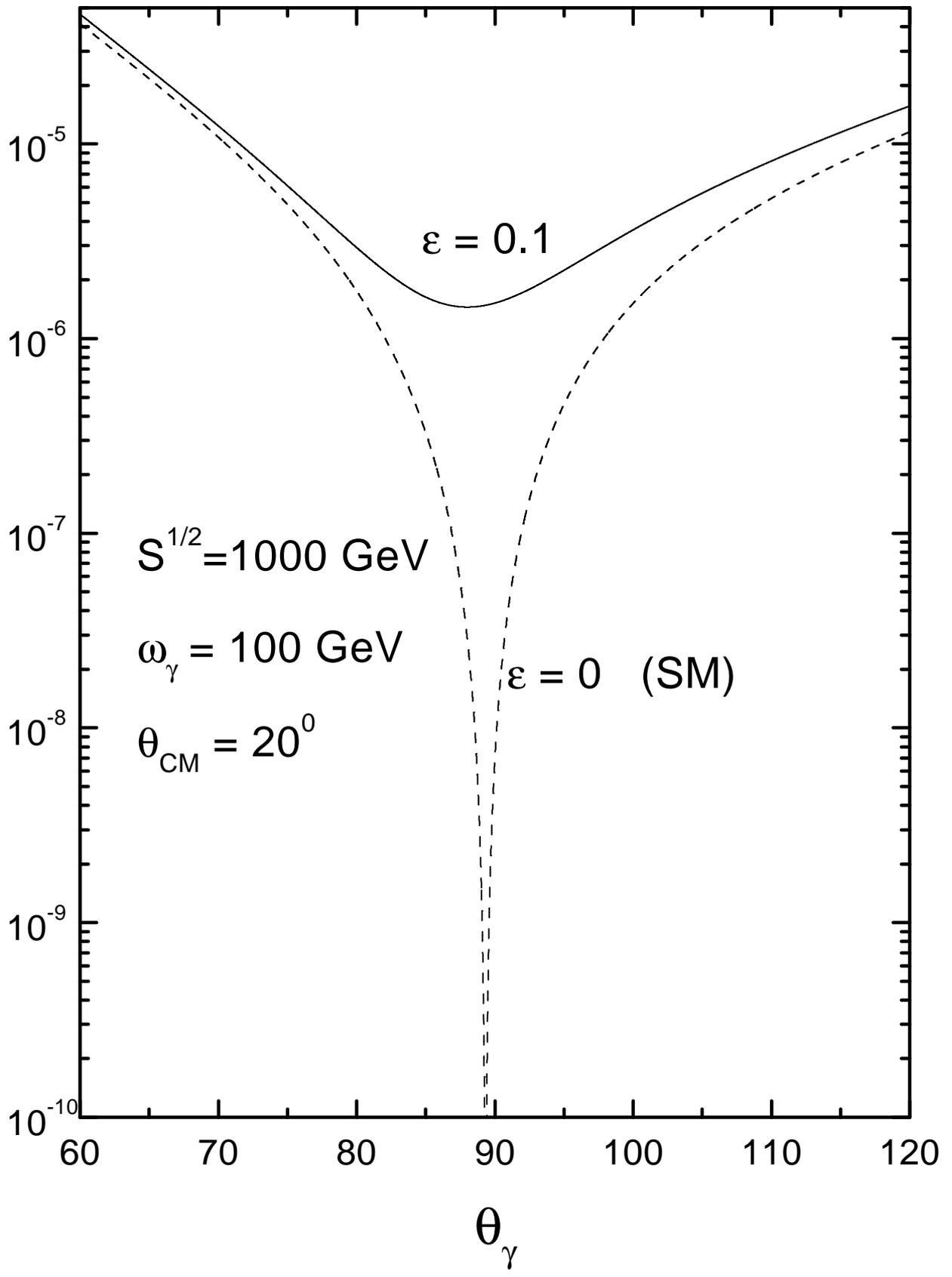
\noindent{\bf Figure 4:} Amplitude squared, in an arbitrary unity,
for the standard model (dashed line) and for the standard model
plus the anomalous contribution (solid line) for
$\sqrt{s}=1000GeV$, $\omega_{\gamma}=100GeV$, $\theta_{CM}=20^0$
and $\epsilon_Z=\epsilon_{\gamma}=0.1$

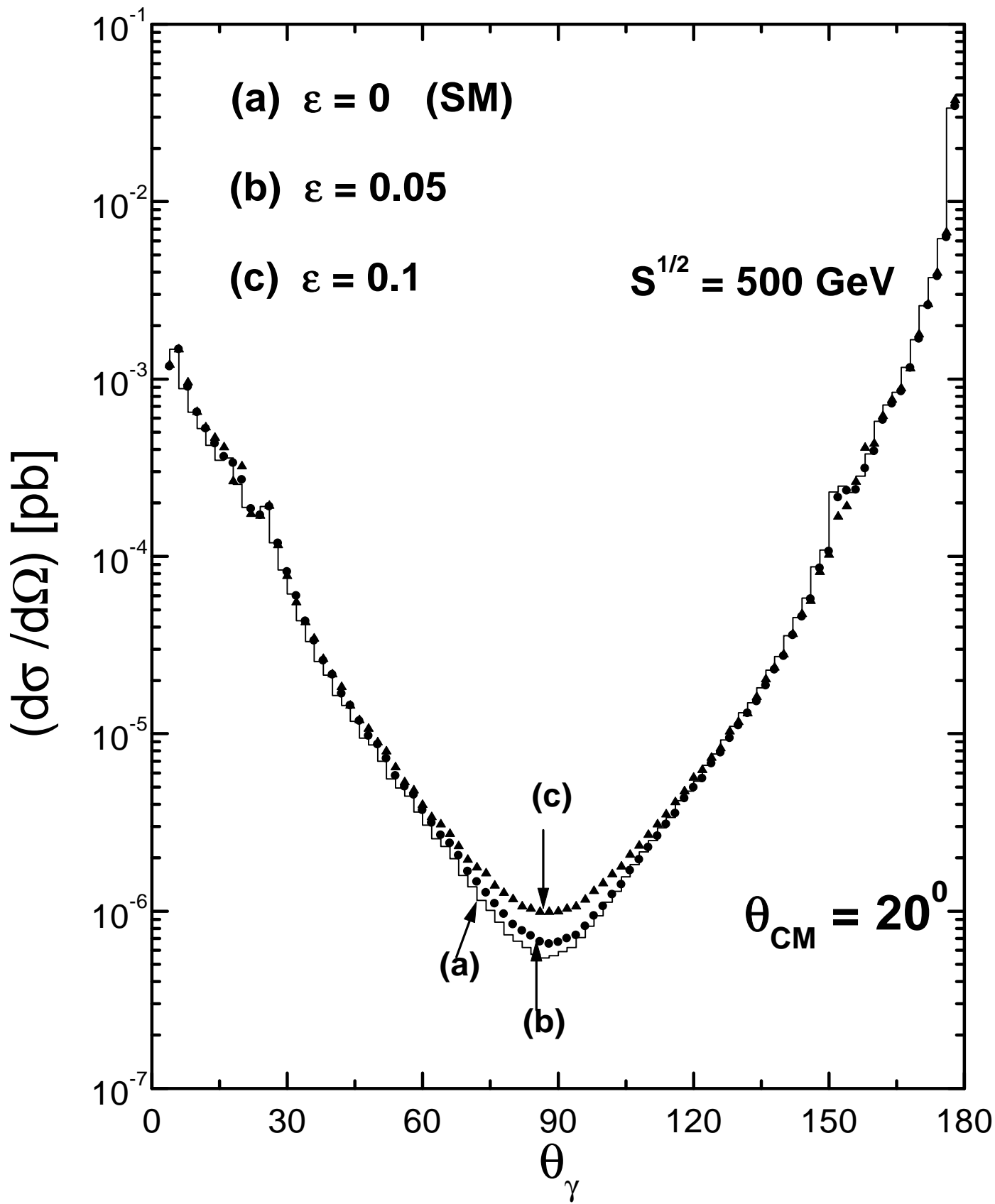
\noindent{\bf Figure 5:}Differential cross-section for the $e^+e^-
\rightarrow \tau \bar\tau \gamma$ scattering as a function of
$\theta_{\gamma}$ for $\theta_{CM}=20^0$ and $\sqrt{s}=500GeV$.
Solid step line represents the standard model contribution
($\epsilon_{\gamma}=\epsilon_Z=\epsilon=0$). Up triangle and dark
circle represent the standard model plus the anomalous
contribution for $\epsilon_{\gamma}=\epsilon_Z=\epsilon=0.1$ and
$0.05$ respectively.

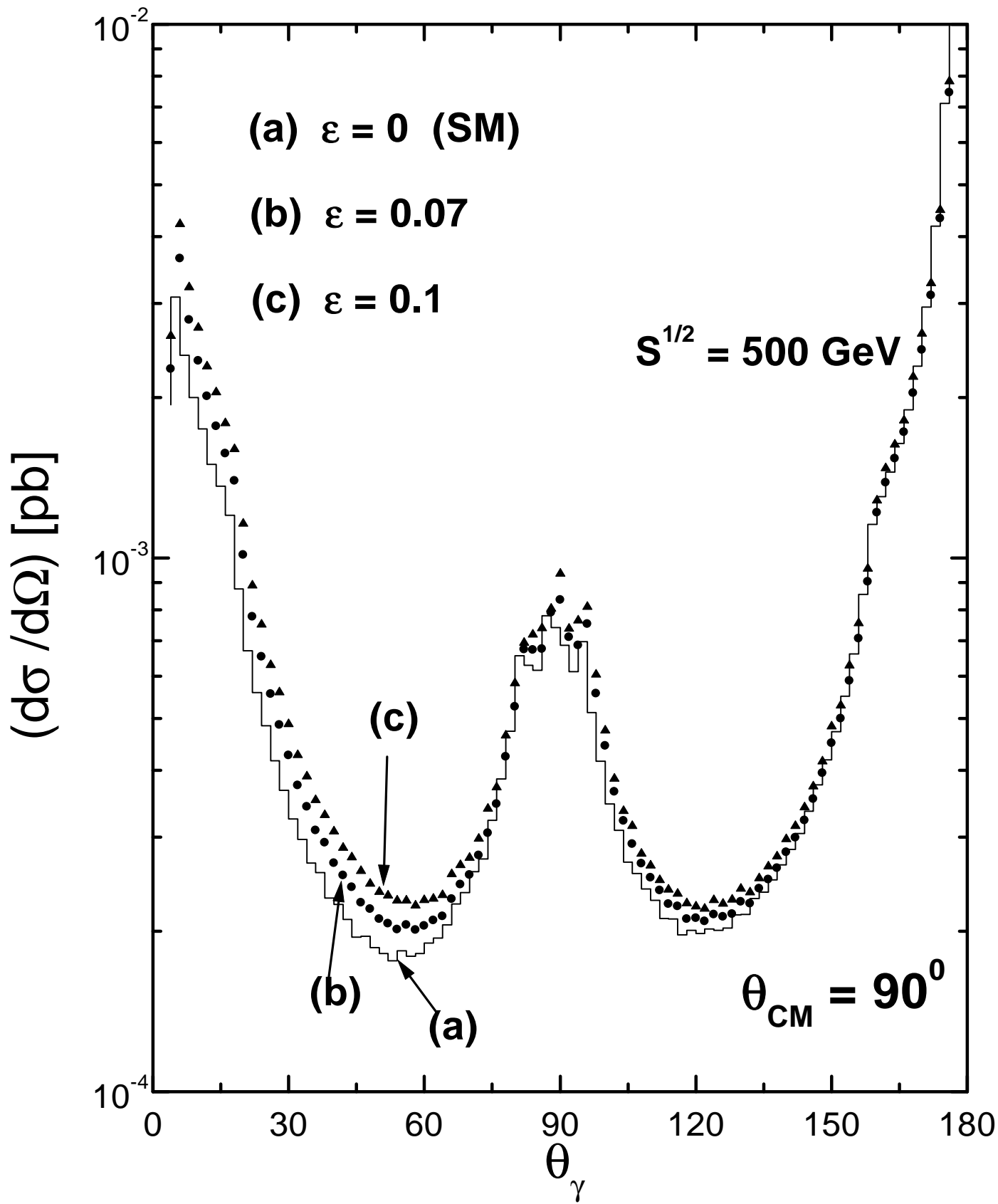
\noindent{\bf Figure 6:} Differential cross-section for the
$e^+e^- \rightarrow \tau \bar\tau \gamma$ scattering as a function
of $\theta_{\gamma}$ for $\theta_{CM}=90^0$ and $\sqrt{s}=500GeV$.
Solid step line represent the standard model contribution
($\epsilon_{\gamma}=\epsilon_Z=\epsilon=0$). Up triangle and dark
circle represents the standard model plus the anomalous
contribution for $\epsilon_{\gamma}=\epsilon_Z=\epsilon=0.1$ and
$0.07$ respectively.

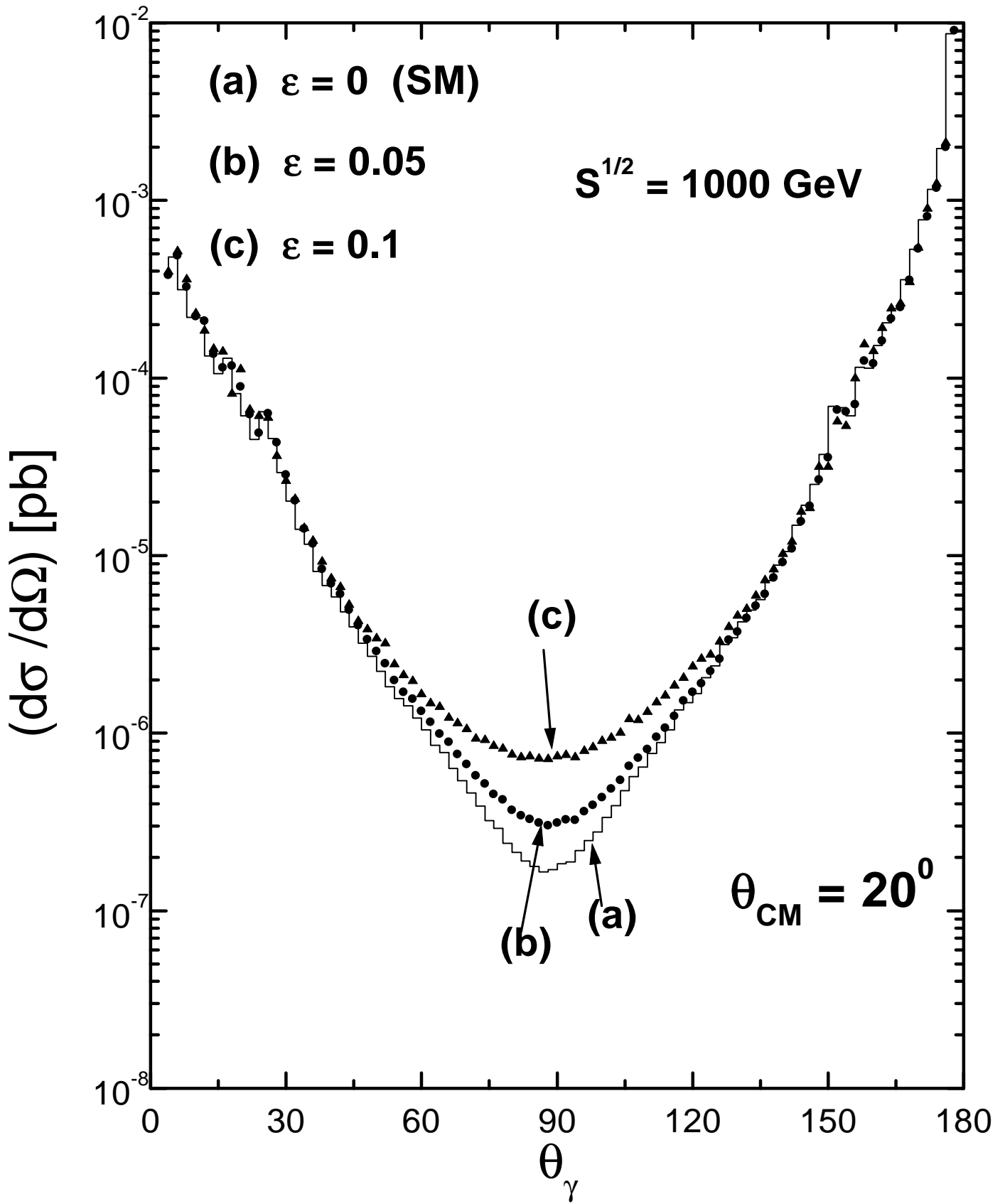
\noindent{\bf Figure 7:} Differential cross-section for the
$e^+e^- \rightarrow \tau \bar\tau \gamma$ scattering as a function
of $\theta_{\gamma}$ for $\theta_{CM}=20^0$ and
$\sqrt{s}=1000GeV$. Solid step line represent the standard model
contribution ($\epsilon_{\gamma}=\epsilon_Z=\epsilon=0$). Up
triangle and dark circle represents the standard model plus the
anomalous contribution for
$\epsilon_{\gamma}=\epsilon_Z=\epsilon=0.1$ and $0.03$
respectively.

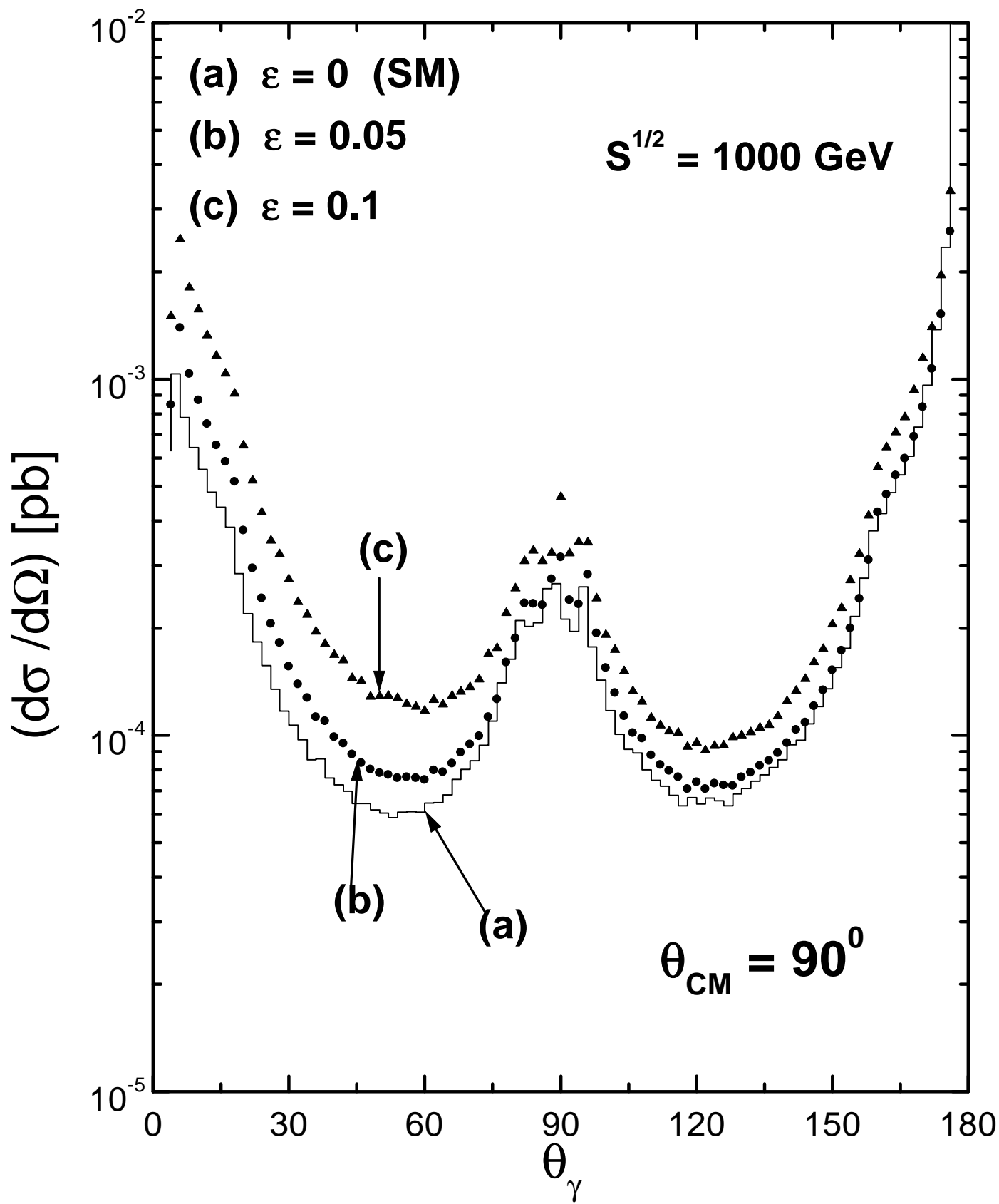
\noindent{\bf Figure 8:} Differential cross-section for the
$e^+e^- \rightarrow \tau \bar\tau \gamma$ scattering as a function
of $\theta_{\gamma}$ for $\theta_{CM}=90^0$ and
$\sqrt{s}=1000GeV$. Solid step line represent the standard model
contribution ($\epsilon_{\gamma}=\epsilon_Z=\epsilon=0$). Up
triangle and dark circle represents the standard model plus the
anomalous contribution for
$\epsilon_{\gamma}=\epsilon_Z=\epsilon=0.1$ and $0.05$
respectively.

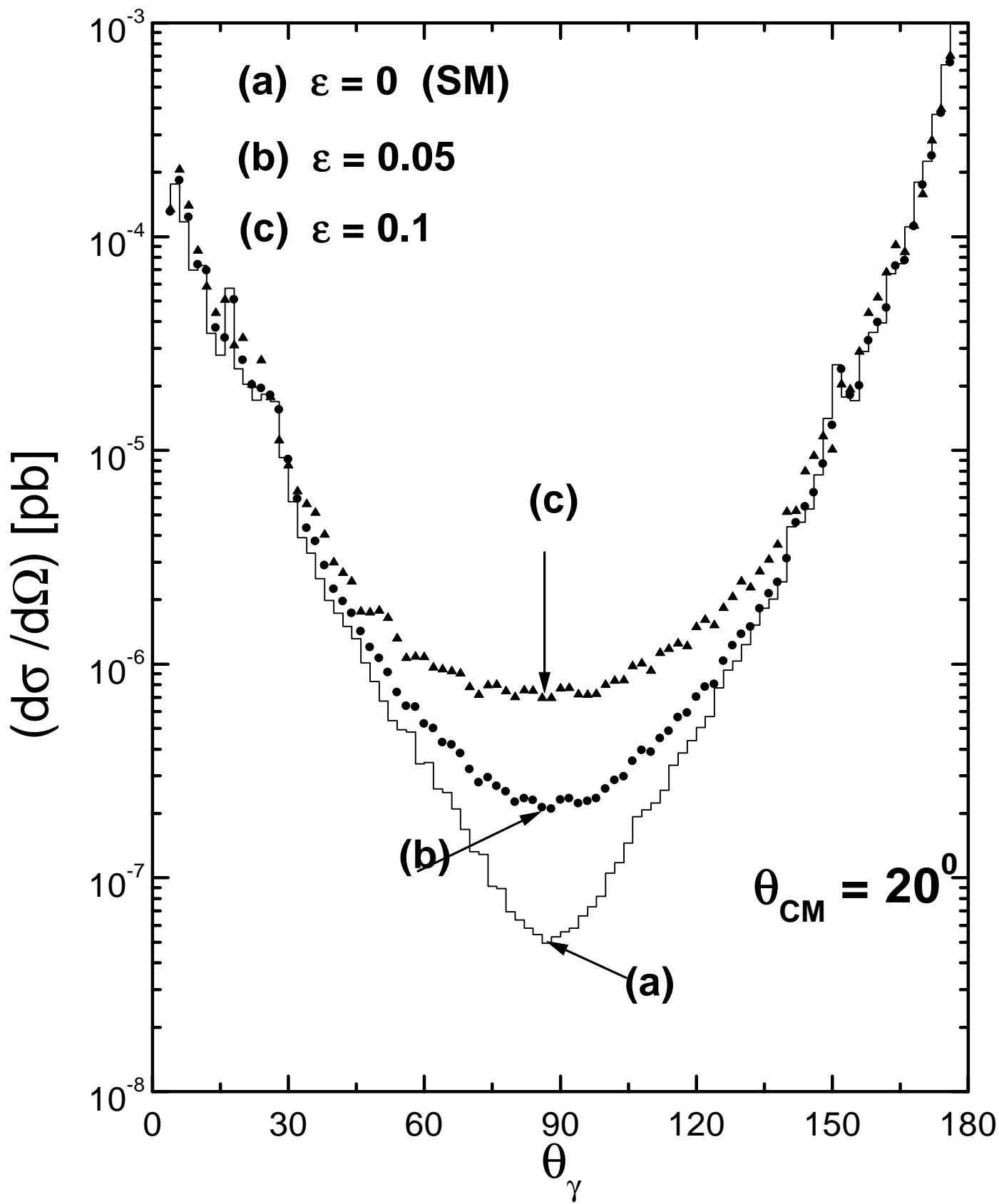
\noindent{\bf Figure 9:} Differential cross-section for the
$e^+e^- \rightarrow \tau \bar\tau \gamma$ scattering as a function
of $\theta_{\gamma}$ for $\theta_{CM}=20^0$ and
$\sqrt{s}=2000GeV$. Solid step line represents the standard model
contribution ($\epsilon_{\gamma}=\epsilon_Z=\epsilon=0$). Up
triangle and dark circle represent the standard model plus the
anomalous contribution for
$\epsilon_{\gamma}=\epsilon_Z=\epsilon=0.1$ and $0.05$
respectively.

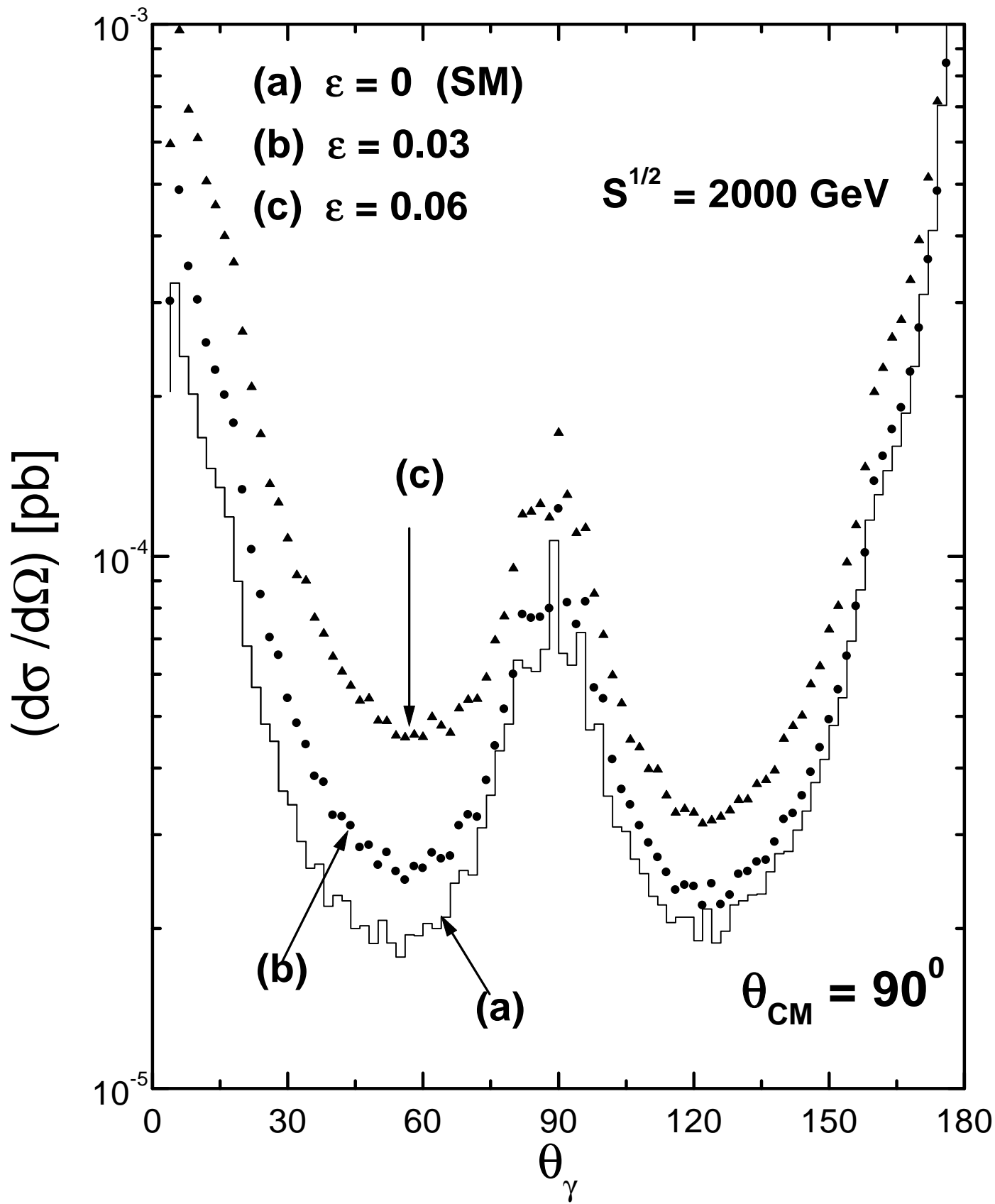
\noindent{\bf Figure 10:} Differential cross-section for the
$e^+e^- \rightarrow \tau \bar\tau \gamma$ scattering as a function
of $\theta_{\gamma}$ for $\theta_{CM}=90^0$ and
$\sqrt{s}=2000GeV$. Solid step line represent the standard model
contribution ($\epsilon_{\gamma}=\epsilon_Z=\epsilon=0$). Up
triangle and dark circle represents the standard model plus the
anomalous contribution for
$\epsilon_{\gamma}=\epsilon_Z=\epsilon=0.06$ and $0.03$
respectively.

\end{document}